# A Data-driven Latent Semantic Analysis for Automatic Text Summarization using LDA Topic Modelling

Daniel F. O. Onah
*Department of Information Studies*
*University College London*
London, United Kingdom
d.onah@ucl.ac.uk

Elaine L. L. Pang
*Academic Skills Development*
*Brunel University London*
London, United Kingdom
elaine.pang@brunel.ac.uk

Mahmoud El-Haj
*School of Computing and Communications*
*Lancaster University*
Lancaster, United Kingdom
m.el-haj@lancaster.ac.uk

*Abstract*—With the advent and popularity of big data mining and huge text analysis in modern times, automated text summarization became prominent for extracting and retrieving important information from documents. This research investigates aspects of automatic text summarization from the perspectives of single and multiple documents. Summarization is a task of condensing huge text articles into short, summarized versions. The text is reduced in size for summarization purpose but preserving key vital information and retaining the meaning of the original document. This study presents the Latent Dirichlet Allocation ($LDA$) approach used to perform topic modelling from summarised medical science journal articles with topics related to genes and diseases. In this study, $PyLDAvis$ webbased interactive visualization tool was used to visualise the selected topics. The visualisation provides an overarching view of the main topics while allowing and attributing deep meaning to the prevalence individual topic. This study presents a novel approach to summarization of single and multiple documents. The results suggest the terms ranked purely by considering their probability of the topic prevalence within the processed document using extractive summarization technique. $PyLDAvis$ visualization describes the flexibility of exploring the terms of the topics' association to the fitted LDA model. The topic modelling result shows prevalence within topics 1 and 2. This association reveals that there is similarity between the terms in topic 1 and 2 in this study. The efficacy of the $LDA$ and the extractive summarization methods were measured using Latent Semantic Analysis ($LSA$) and Recall-Oriented Understudy for Gisting Evaluation ($ROUGE$) metrics to evaluate the reliability and validity of the model.

## I. Introduction

Topic modelling has been performed on several types of documents in the past. However, this study presents a novel approach to topic modelling by performing extractive summarization on over 100 articles related to genes and associated diseases and feeding the summary as an input argument a Latent Dirichlet Allocation (LDA) model in order to perform the topic modelling. The idea here is to identify the commonalities between articles of the same genre describing a specific topic of interest in the research. The study is addressing journal articles retrieved from PubMed Central (PMC[1]) database discussing about genes and their associated diseases.

What would you do if you were handed a pile of papers—receipts, emails, travel itineraries, meeting minutes—and asked to summarize their contents? One strategy might be to read through each of the documents, highlighting the terms or phrases most relevant to each, and then sort them all into piles. If one pile started getting too big, you might split it into two smaller piles. Once you had gone through all the documents and grouped them, you could examine each pile more closely. Perhaps you would use the main phrases or words from each pile to write up the summaries and give each a unique name—the topic of the pile. This is, in fact, a task practiced in many disciplines, from medicine to law, from computer science to engineering and so on. At its core, this sorting task relies on our ability to compare two documents and determine their similarity. Documents that are similar to each other are grouped together and the resulting groups broadly describe the overall themes, topics, and patterns inside the corpus. With so many documents being extracted from social media, review comments from online platforms and microblogs as Twitter, a huge amount of natural language data is being mined and are available to be analysed [1]. Certainly, it is reasonable to translate summarized documents accurately. An example, articles extracted in a different language from English to be translated to make sense similar to the original language of which the article was written [2].

Automatic text summarization is the process of performing specific NLP task by producing a concise summary of documents (single or multiple) without any manual support while preserving the meaning or important points of the original document [3]. In this study, we try to answer the following research questions:

- How automated text summarization techniques were used in an extractive summary of articles?

[1] https://www.ncbi.nlm.nih.gov/pmc/

- How topic modelling models were used in producing emerging terms that are related to multiple and different journal articles?
- Are the search terms used for the text mining of articles from the database predominant in the emerging terms that were extracted from the processed text?

The experimental results show that the proposed model achieves good performance in terms of the document summary and the topic modelling information retrieved from the full document. This paper is presented as follows. Section 2, covers related study on summarization. Section 3, conceptualise the methods and describes the techniques applied in the study. Section 4 describes topic modelling, Section 5, presents a description of the model pipeline. Section 6, presents the results and findings. The limitation of the study is presented in section 7 and finally, the discussion and conclusion of the study are presented in sections 8 and 9.

## II. Related Work

The amount of text data being produced worldwide is enormous and growing rapidly. Unless these text data are extracted and make meaning, then the most important and relevant information would be lost. Text summarization is a well-known task in natural language understanding and processing. Summarization is described as the process of presenting huge data information in a concise manner while focusing on the most useful sections of the data whilst preserving the original meaning [4]. The most important element of text summarization is to produce a clear and concise summary taken from the large datasets that would make sense to the reader and direct to the main points [5]. There is a need for automation of these increasingly available web text data for information retrieval and sustainability. In this modern era of big data, text mining has been retrieved from various sources, website, databases, journals and conference articles in related studies. The voluminous text data need to be collected and summarised in order to retrieve useful information concerning the main content of the document.

### A. Summarization

Summarization is a technique in NLP that is used for condensing or summarising huge texts into smaller versions taking care not to omit the main relevant information contained in the document [6]. This helps in reducing the size of the original document either single or multiple while preserving key elements and meaning of the content [7]. This is the main significant of automatic summarization, by presenting the documents in a more meaningful manner. Manual summarization is tedious, expensive and laborious to under-take [8]. There is need for automated summarization which is gaining popularity among researchers. There are so many important models for performing automatic text summarization in various NLP tasks such as classification, automatic question and answering, computational journalism, financial summarization, news summarization and foreign language summary translation. One of the key factors of the document summary is that it can be integrated into these NLP applications to reduce the size of the document for processing while possibly retaining the original information contained in the document [9], [10]. There are two different approaches to automatic summarization; these are extraction and abstraction.

*1) Extractive approach:* Extractive summarization approach considers the top *N* sentences based on their score rankings for the summary generation [11], [12]. This paper focuses on extractive text summarization. The study focuses on direct object extraction from the original document without any modification of the content. Extractive summarization approach takes object as input and generate the summary based on the probability vector [4]. Word frequencies are considered as one of the input factors in the sentence score rankings which represent the probability of the sentence to be included in the summary. In order to generate the final summary, the best sentence scores are selected based on the maximum number of words in the sentence and the number of sentences that met the specified threshold provided [13]. We will briefly describe the abstractive approach of summarization and explain why we decided to use extractive approach in this study.

*2) Abstractive approach:* In abstractive text summarization technique, this follows the convention of unsupervised approach where machine learning paradigms such as deep learning plays a big role in generating the document summary [14]. This approach considers a bottom-up summary for which some of the sentences might not be part of the original document [15]. However, in some cases, the vocabulary of the documents might be the same as the original document [16]. Designing an abstractive model for summarization is very problematic and challenging because it involves a more complex language modelling [17].

In this study, we decided to use extractive approach for article summarization, because we wanted all parts of the sentences that will be summarised to be from the original document.

### B. Topic Modelling

Topic modelling is the process of labelling and describing documents into topics. This is an unsupervised machine learning technique for abstracting topics from collections of documents [18]. Topic modelling approach is based on an inductive modelling used to abstract core themes from a weighted graphical representation of documents obtained during the processing stages. In order to apply topic models in NLP application, there is need for extrapolation of topics from unstructured datasets. In this study, $Scitkit - Learn$ and $Gensim$ were used to extract the topics from the models using $^{0}gensim.models.ldamodel.LdaModel^{0}$, which takes in as input argument the text corpus, number of topics to be extracted and $id2word$ that contains the dictionary terms for the preprocessed document [19].

### C. Latent Dirichlet Allocation

Latent Dirichlet Allocation (LDA) is a technique applied in topic modelling introduced by [20]. This is a topic discovery technique used to generate topics based on the probability that each given term might occur within the document. The document can be in the form of mixture of topics that might not necessarily be distinct and words may appear in multiple topics [18], [21]. In this approach, presented with words or token from multiple documents from which a probability topics model is constructed, we observed word distributions for each mixture of topics in the document.

## III. Methods

The summarization model was designed to scrap text data from PubMed journal database using genes and diseases keywords search [22]. A web-scraping model that was used to retrieved the articles for this research was able to scrape about 100 papers at a time from the PubMed Central (PMC) repository. We could have extracted more articles. However, we wanted to use this sample as the initial based study. The model applied some NLP techniques for the initial preprocessing of the data for extractive summarization. The proposed model in this study is scalable and generalizable for producing arbitrarily size summaries by splitting the documents into resemble content. The study applied sentence scoring on the clean document to extract text that fell within the threshold of high frequency score used in the model. During the summarization process, we calculated word frequency and the high sentence scores that was used to summarise the articles. We created vectors to store the sentences. This allows us to fetch summary for 100 elements for the constituent words in a sentence. Finally, we took the mean of those vectors to consolidate the vector for the sentence. The next phase is to perform a cosine similarity scores of the sentences using a matrix dimensions of $n * n$, where $n$ is the number of sentences in the document. The cosine similarity was applied to perform the similarity between a pair of sentences. We then extract the top $N$ sentences based on their ranking for the summary generation that was then fed into the LDA topic modelling.

The study was designed to apply a generalised concept of LDA topic modelling technique to create a dictionary of terms that was fed from the summarised articles. This dictionary of terms was used to build a vectorised corpus of lexicon LDA model. One of the key approaches that was used in the experiment was the *'pyLDAvis.gensim.prepare'* method which takes as an argument our LDA model, the corpus and the derived lexicon which contains the dictionary terms for the study [23], [18], [24], [19]. Another method that was in the study was the '$gensim.models.ldamodel.LdaModel^0$, which takes the summary corpus as an input argument, the number of topics to be extracted and the '$id2word^0$ that contains our dictionary terms for the journal articles. This method allows us to project the topics by calling the method that help in visualising the interactive topic modelling shown in Figures 7 and 8.

### A. Model Description

Most of the processing of the text was performed with the python Natural Language Toolkit (NLTK) [25], including using the NLTK Tokenizer to tokenize the text. The overarching research model was developed to retrieve specific information from huge published journals using the topic modelling approach of NLP. In this study, we used multiple journal articles related to diseases and genes with sequence of paragraphs. The task is to generate the summary at most predominant sentence level. Extractive summarization approach applied in the study produced naturally grammatical summaries without much linguistic connotation or analysis. Since extractive summarization uses a supervised technique, the sentence selection process involves scoring each sentence in the original cleaned document. In this case, a label is produced to indicate whether a sentence met the conditions which are the chosen length of the sentence or the summary threshold indicated in the model. It is only when these predefined conditions are met before a sentence could be considered to be included in the final summary. The supervised learning method allows for maximization of the likelihood of sentence consideration from the input document. This approach could also be generalised on other articles such as media, blogs, news and so on and will produce the same outcomes.

*1) Sentence scoring method: :* In this study, a scoring function is introduced to generate the sentence score dictionary which hold the value assigned to each sentence [26]. This denotes the probability that the sentence will be selected and included in the summary (see Figure 1). The summary length is fixed, therefore, the top $N$ sentences with the highest score rankings are chosen for the summary. In this study, the quality of the document summary largely depends on the chosen sentences and this would reveal the relevance of the information retrieved from within the full document. The process of scoring the sentence is represented in equations 1 and 2. If the sentence is not in the sentence score dictionary keys during the processing, the words in the word frequencies dictionary is added to the sentence scores (see equation 1 ).

$$\mathrm{Sent}_{scores}[S] = Word_{freq}[W] \tag{1}$$

During the sentence model processing interval, the length of the sentence is either increased or reduced by certain values within the sentence scores dictionary. Therefore, new sentences are added into the sentence dictionary scores. The sentence model would check whether the new sentences are in the sentence dictionary. If the sentence exists in the sentence dictionary, then the model will proceed accordingly. But if the process sentence is not in the sentence scores dictionary keys, then the word in the word frequencies dictionary is added to the sentence in the sentence scores dictionary (see equation 2).

$$\text{Sent}_{scores}[S] += Word_{freq}[W] \qquad (2)$$

*2) Word frequency::* Dictionary of word frequency corpus was generated within the model. The word frequencies were selected automatically based on the prevalence or occurrence of the words in the corpus dictionary created in the model (see Figure 2). The length of the sentence selected for the word frequencies was less than 30. Sentences with less than

```
, 'Furtherit can be used to identify approved drugs that can be repurposed for c
ardiac disease treatments.': 1.4402730375426622, 'Specificallydrugs that are app
roved for non-cardiac therapiestarget high-ranking Hridaya-genesand are upregula
ted in DCM patients should be considered top targets for cardiac-drug repurposin
g.': 1.3856655290102389, 'Hridaya can also predict genes having significantly di
fferent exon usage in DCM patient heart.': 2.0477815699658706, 'Stratifying DCM
patientsusing either the expression or the genetic regulators of predicted funct
ional genesreveal two distinct subgroups of patients with different clinical phe
notypes.': 2.0477815699658706, 'Due to various confoundersespecially co-expressi
on among genesthe clear majority of differential genes are likely to represent d
ownstream effects.': 2.054607508532423, 'In comparisonamong the top 1000 Hridaya
-genes 84% are differentially expressed.': 1.416382252559727, 'Interestinglyhowe
verthe clear majority of these genes (76%) are down regulated in DCM individuals
.': 3.5255972696245736, 'Though importantanimal models have been found in many c
ases to have poor translatability47.': 1.0853242320819112, 'The previous studies
 did not investigate the genetic signals underlying gene expression to detect fu
nctional genes.': 2.696245733788396, 'In contrastour approach integrates a wide
range of geneticepigenetictranscriptomicphenotypicand evolutionary evidence and
utilize data from 213 human hearts to predict functional genes of DCM.': 2.34470
98976109215, 'Several previous works have directly addressed this broad problemb
oth regarding identifying functional genes as well as associated SNPs underlying
 diseases.': 2.225255972696246, 'As an examplea framework called Combined Annota
tion Dependent Depletion (CADD) uses SVM to estimate relative pathogenicity of h
uman SNPs49.': 2.8600682593856654, 'Polyphen-2 predicts damaging effects of miss
```

Fig. 1. Tokenize sentence score for the article summary

30 ($< 30$) words were selected. These maximum weighted frequency ($Freq_{max}$) of each word were calculated by using the product of the word frequencies ($W_{freq}$) and the values ($V$). These are then added to final summary (see equation 3).

$$\text{Freq}_{max} = Max(W_{freq} * V) \qquad (3)$$

The next equation allows us to calculate the maximum word in the word frequencies (see equation 4).

$$\text{Word}_{freq}[W] = \frac{Word_{freq}[W]}{Freq_{max}} \qquad (4)$$

```
dict_keys(['Warning', ':', '\\tThe', 'NCBI', 'web', 'site', 'requires', 'JavaScr
ipt', 'function', '.', '\\tmore', '...', '1Center', 'Bioinformatics', 'Computati
onal', 'BiologyUniversity', 'MarylandCollege', 'ParkMaryland', '20742', 'USA', '
2The', 'Neufeld', 'Cardiac', 'Research', 'InstituteTel', 'Aviv', 'UniversityTel'
, 'Aviv-YafoIsrael', '3Tamman', 'Cardiovascular', 'InstituteSheba', 'Medical', '
CenterRamat', 'GanIsrael', '4The', 'Dr.', 'Pinchas', 'Borenstein', 'Talpiot', 'L
eadership', 'ProgramSheba', 'CenterTel-HashomerIsrael', '5Department', 'Radiatio
n', 'OncologySheba', '6The', 'Blavatnik', 'School', 'Computer', 'ScienceTel', 'A
viv69978', 'Israel', 'Idiopathic', 'dilated', 'cardiomyopathy', '(', 'DCM', ')',
 'complex', 'disorder', 'genetic', 'environmental', 'component', 'involving', 'm
ultiple', 'genesmany', 'yet', 'discovered', 'We', 'integrate', 'geneticepigeneti
ctranscriptomicphenotypicand', 'evolutionary', 'features', 'method', '-', 'Hrida
yato', 'infer', 'putative', 'functional', 'genes', 'underlying', 'genome-wide',
'fashionusing', '213', 'human', 'heart', 'genomes', 'transcriptomes', 'Many', 'i
dentified', 'Hridaya', 'experimentally', 'shown', 'cause', 'cardiac', 'complicat
ions', 'validate', 'top', 'predicted', 'genesvia', 'five', 'different', 'analyse
s', 'Firstthe', 'associated', 'cardiovascular', 'functions', 'Secondtheir', 'kno
ckdowns', 'mice', 'induce', 'abnormalities', 'Thirdtheir', 'inhibition', 'drugs'
, 'side', 'effects', 'Fourththey', 'tend', 'differential', 'exon', 'usage', 'nor
mal', 'samples', 'Fifthanalyzing', 'individual', 'genotypeswe', 'show', 'regulat
ory', 'polymorphisms', 'elevated', 'risk', 'The', 'stratification', 'patients',
'based', 'expression', 'reveals', 'two', 'subgroups', 'differing', 'key', 'pheno
```

Fig. 2. Dictionary of word frequency corpus

*B. Research Pipeline*

The pipeline model for the research follows a sequential approach of processes that could allow the smooth and efficient information retrieval. The pipeline in Figure 3 was used to answer the research questions in this study.

*1) Data Collection:* The dataset was scraped from the web. About 100 papers were extracted from PubMed Central (PMC) database ("*https : //www.ncbi.nlm.nih.gov/pmc/*"") using a search key combination of *'gene'* and *'disease'*. The articles scraped from the web were all related to medical science research. Papers related to diseases and the mutated genes causation were extracted for this study [22]. These papers were extracted with HTML tags that are required to be preprocessed, cleaned and summarized for the topic modelling approach.

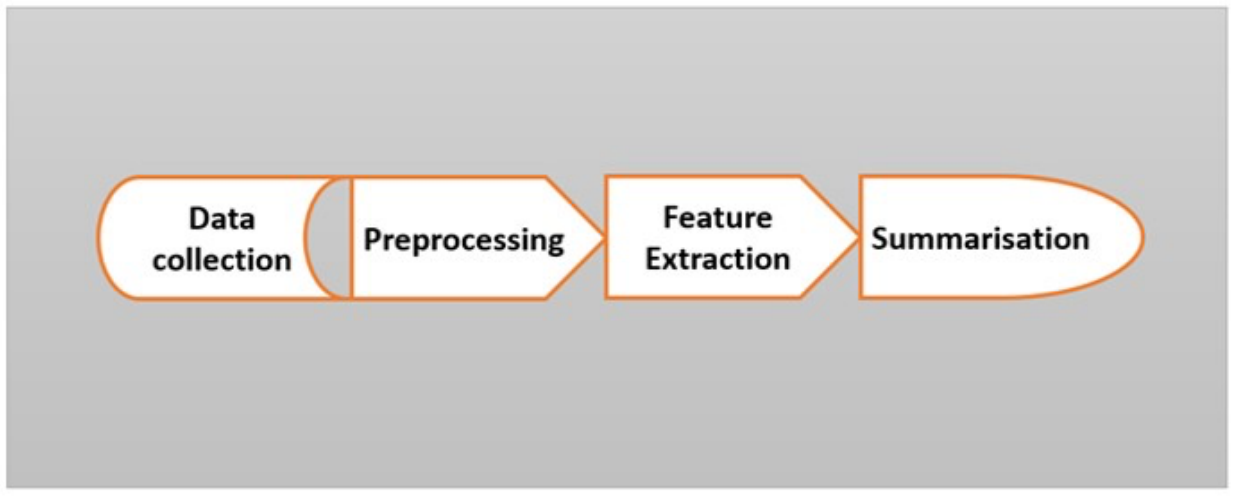

Fig. 3. Pipeline of text mining processing

*2) Pre-processing & Feature Extraction:* The web-based dataset scraped from PubMed journal was in raw state and unstructured which consist of HTML tags, special characters, symbols and numbers that had to be processed and cleaned. The preprocessing involved converting the dataset into text documents using NLP packages such as BeautifulSoup, regular expression, lxml, tokenisation and using NLTK library. In the feature extraction process, we parse the web articles source code in order to extract the textual material needed for the final summary. As the articles were parsed through the source code, the text for extraction are between the paragraphs' tags $< p >$ *text* $< /p >$. During the process of formatting the clean articles, we performed extra filtering of special characters from the processed text in order to find and replace these symbols automatically. Finally, these extracted paragraphs text are combined to form a single string to store the clean web content for further topic model processing (see Figure 4).

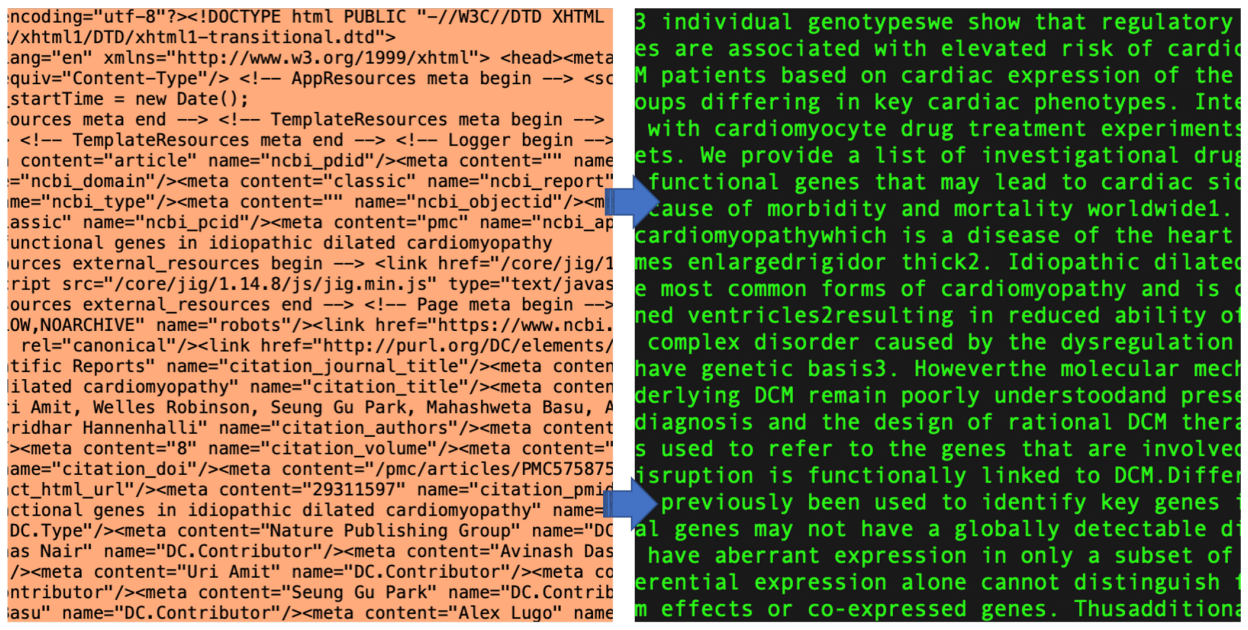

Fig. 4. Raw HTML article dataset processed to clean text

*3) Stopwords:* We further removed a list of stop-words from the propocessed articles. Words such as pronounce that are not necessary or essential for the final summary (see Figure 5 for the list of stop words).

*4) Topic Modelling & Visualization:* This study was able to reveal prevalence of terms that emerged within the documents and show their relevance by how the projection of the topic modelling circle and the size of a word in the result visualisation. The result was visualised using PyLDAvis which is a web-based interactive visualization package that allows the display of the topics that were identified using the LDA approach [27]. PyLDAvis was used for extracting information from the fitted LDA topic models to design a web-based interactive visualization. The main method that was applied in this study was *'pyLDAvis.gensim.prepare'* which takes as an argument topic models from LDA, the vectorized text corpus, and the derived lexicon which contains the dictionary terms from the study [18], [24]. Each identified topic is encoded in the circles of the PyLDAvis and the bigger the circle the more projection or prevalence is the topic (as seen in Figures 7 and 8). The higher the number of common words among sentences indicates that the sentences are semantically related.

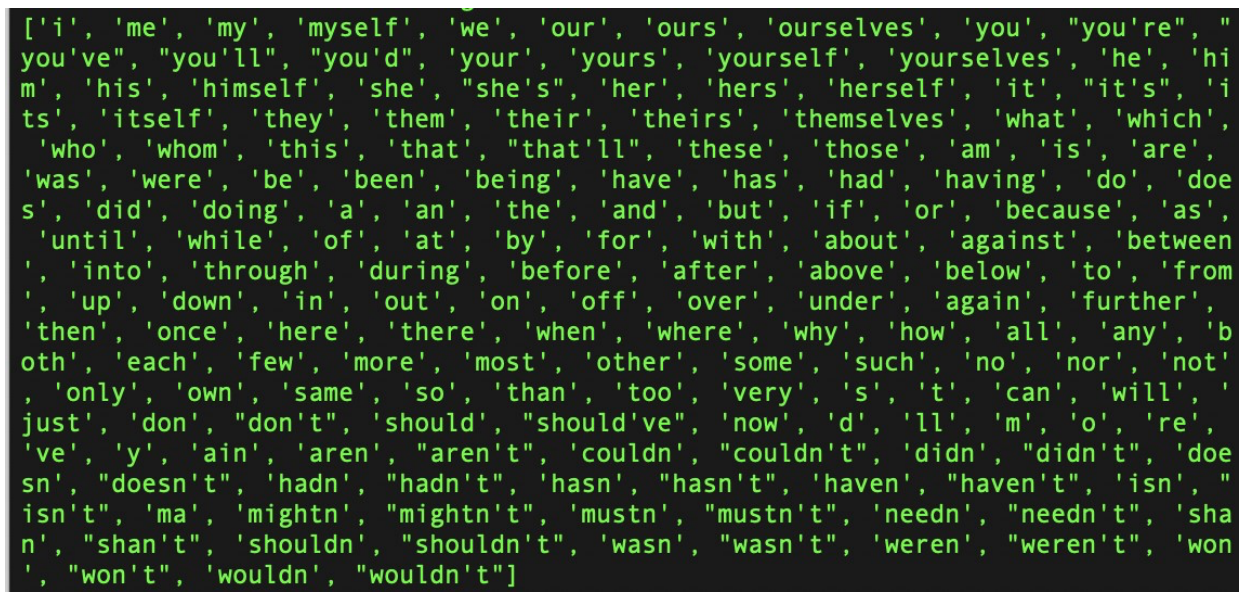

Fig. 5. List of stop-words removed from the article

## IV. Model

*A. Defining semantic significance* we define the semantic significance of term *t* to the topic

*n* given the parameter weight of the $(\lambda)$ $where(0 \leq \lambda \leq 1)$ [24]. Let *pt* denotes the minimal probability of the term *t* in the lexical corpus. Let *nt* denote the probability of term *t* element of *1, ..., N* for n element of *1,..., K*, where *N* denotes the frequency of terms in the vocabulary (see equation 1)

$$s(t,p|(\lambda)\ ) \ = \ \sum \lambda \ log(\Omega_{nt}) + (1-\lambda\ )log\frac{(\Omega_{nt})}{pt} \quad (5)$$

where $(\lambda)$ is the weight given to probability of the terms *t* in topic *n* (equation 1)

*B. Defining Saliency Term*

In this study we define saliency term as given a word '$w^0$, we compute its minimal probability *P(TM/w)*. where *TM* is the topic model. The possibility that the emerge word *w* was generated from the *LDA* topic model *(TM)*.

We also compute the marginal probability *P(TM)*: - with the possibility that any word $w^0$ randomly selected was generated by *TM*. We define the uniqueness of each identified word $^0w^0$ as the divergence occurrence between *P(TM/w)* and *P(TM)* [28]: we were able to compute 5 topics *(t)* and 10 passes which were selected from the Latent Dirichlet Allocation (LDA) topic modelling (see equation 6).

$$\text{U(6)} \quad _w = \sum_{t}^{10} P(TM_t/w)log\frac{P(TM_t/w)}{P(TM_t)} \quad _{=5}$$

The uniqueness of each term is described as how significance and semantically associated they are to the topics. For example, a term could be semantically associated to more than one topic. The frequency and population of terms are denoted by the size of the topic circles and also the inter-topic distance denote how closely related the topics are. We notice a few words that are expressed in several topics, but observing this word *w* reveals little information about the mixture or semantic association of the topics. In some cases, this word might be scored very low in the computation of it's uniqueness. In order to compute the saliency, we used the following model equation 7:

$$S_w = P(w) * U_w \quad (7)$$

As illustrated in Figure 8, adjusting the lambda metric can aid in the significant classification and reducing the complexity of the topics. This helps to remove ambiguity of the terms association by making term distribution clearly. Looking at the figure, we observed that given equal frequency of words, the list of the most common, relevant or distinctive terms (e.g. *gene*, *disease*, *expression*, *associate* ) are prevalence in the visualised graph-plot distribution. The saliency measures the distribution of the speeds and identification of topic association and composition (e.g. prevalence topic 1 terms such as *genes,disease* etc. These terms are all semantically associated to topics 2 and 3).

## V. Latent Semantic Analysis

Latent Semantic Analysis (LSA) is a robust Algebraic and Statistical method which extracts hidden semantic structures of words and sentences. LSA is used to extract features that cannot be directly mentioned within the dataset [29]. These features are essential to data, but are not original features of the dataset. It is an unsupervised approach along with the usage of Natural Language Processing (*NLP*). It is an efficient technique in order to abstract out the hidden context of the document [30]. We performed a mini summary from the original summary from the study using latent semantic analysis (LSA) for text summarization. The mini summary was done from the summary of the original clean 100 articles extracted from *PubMed* (*https* : *//pubmed.ncbi.nlm.nih.gov/*) database. This summary is then fitted into the ROUGE metric system to measure the efficacy of the model. Results from the LSA present a robust summary of the entire articles with useful information extracted about specific genes that are associated to cancer disease. Below is the summary and visualisation of key terms from the summary using a world cloud (Figure 6).

### *A. Sample Extracted Summary*

The sentences with the most prevalence sentence score was used for the summary together. We used the heap queue (*heapq*) library to select the most or very useful sentences. The *heapq* is used in implementing the priority queues for word frequencies in sentences with higher weight is given more priority in processing the summary. The threshold indicates the number of sentences to summarize (see Table III). Different threshold points were selected for the summary and the result indicate differently even though the word frequency selected is less than 30 maximum ($< 30$).

TABLE I

Summary of article using different thresholds

| LSA Extractive Summary | |
|---|---|
| Threshold | Summary |
| >= 3 | 'Some of the genes in the BCAA metabolic pathway such as MLYCD (rank 164)HADHB (rank 354)IVD (rank 713)MUT (rank 921)and PCCB (rank 684) are also ranked highly by Hridaya. The SVMs are based on 181 features broadly grouped into (1) genetic(2) epigenetic(3) transcriptomic(4) phenotypicand (5) evolutionary. The genes are PDGFRBABL1FLT1; and these genes are drug targets of cancer drugs like Dasatinib (targets – PDGFRBABL1)Pazopanib (targets – PDGFRBFLT1)Ponatinib (target – ABL1)26' |
| >= 5 | 'The genes are PDGFRBABL1FLT1; and these genes are drug targets of cancer drugs like Dasatinib (targets – PDGFRBABL1)Pazopanib (targets – PDGFRBFLT1)Ponatinib (target – ABL1)26. For a given genethe product of the two probabilities P'(DCM\|All)2009=˘ 2009˘ P'(Diseasefunctional\|All)xP'(DCM\|Diseasefunctional)called Hridaya-potentialis the final estimated potential of a gene to be a DCM functional gene. Encouraginglywe find that the Hridaya-potentials are much higher for genes having differential exon usage (739 genes) than the rest of the genes (Wilcoxon rank-rump-value2009=˘ 20091.31e-73)˘ .' |
| >= 7 | 'The genes are PDGFRBABL1FLT1; and these genes are drug targets of cancer drugs like Dasatinib (targets – PDGFRBABL1)Pazopanib (targets – PDGFRBFLT1)Ponatinib (target – ABL1)26. For a given genethe product of the two probabilities P'(DCM\|All)2009=˘ 2009˘ P'(Diseasefunctional\|All)xP'(DCM\|Diseasefunctional)called Hridaya-potentialis the final estimated potential of a gene to be a DCM functional gene. Furthermoreas the set of DCM functional genes is a subset of disease functional genes P'(DCM,Diseasefunctional\|All)2009=˘ 2009˘ P'(DCM\|All).' |

### *B. Findings*

The summary result has revealed very interesting findings of genes that are associated to some Cancerous and type 2 diabetes diseases (see Table II).

TABLE II
INFORMATION EXTRACTION

| Diseases & Genes Extracted | | |
|---|---|---|
| Disease | Gene | Drug |
| Cancer disease | PDGFRBABL1FLT1 | Dasatinib |
| Cancer disease | PDGFRBABL1 | Dasatinib |
| Cancer disease | PDGFRBFLT1 | Pazopanib |
| Cancer disease | ABL1 | Ponatinib |
| Dilated Cardiomyopathy | DCM (exon 739 & Wilcoxon) | Hridaya |
| Type 2 Diabetes | BCAA metabolic genes pathway (MLYCD,HADHB,IVD,MUT & PCCB) | Hridaya |

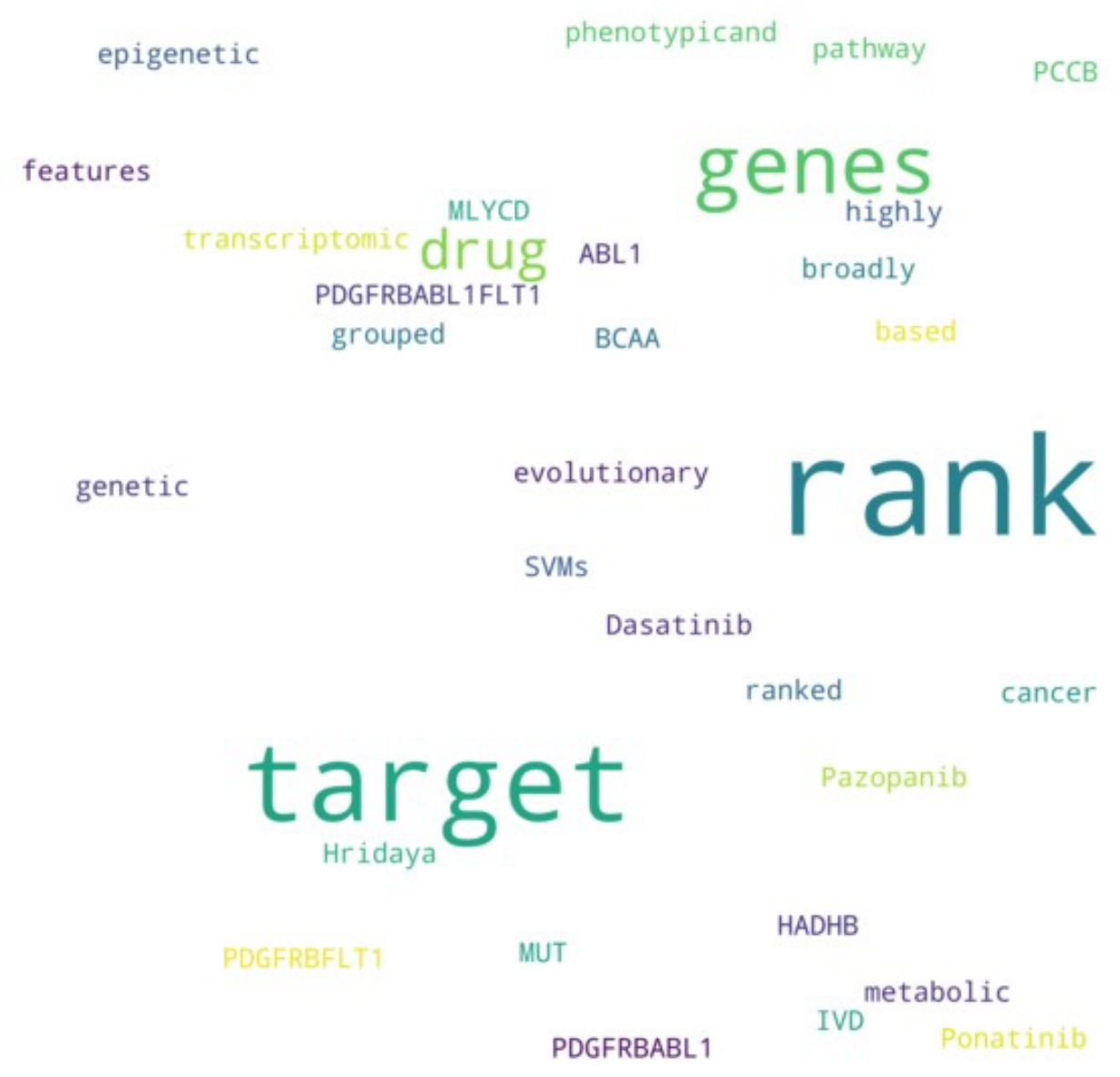

Fig. 6. Word cloud visualisation from the LSA summary (threshold >= 3)

## VI. ROUGE: RELIABILITY & VALIDITY OF MODEL

ROUGE is a metric evaluation model which stands for *Recall Oriented Understudy for Gisting Evaluation*. It is an intrinsic metric for automatically evaluating document summaries [31]. This is originally based on a metric used for machine translation called Bilingual Evaluation Understudy (BLEU). BLEU metric is a score for comparing a machine or candidate translation of text to one or more human annotation or reference translations. Although developed originally for text translation, it can be used to evaluate text generated for a set of natural language processing activities. ROUGE has measures that allow for the evaluation of the accuracy of system summary as compared to a human created summary known as the model summary [32], [33]. The measures were able to count the number of overlapping units of word such as $n - gram$, $bi - gram$ and word pairs between the system generated summaries and the model summaries created by humans. This study introduces a few ROUGE measures: ROUGE-1, ROUGE-2, ROUGE-3, ROUGE-L, ROUGE-S included in the original ROUGE evaluation model and used in this research. ROUGE was used to check for the reliability and validity of our model. After the model is fitted, the external quality of the model is verified according to the fit metric test ROUGE. Common metrics include, but are not limited to, parsimonious fit, valueadded fit, absolute fit and other metrics, and the intrinsic quality of the model is verified through the fit analysis.

TABLE III
ROUGE METRICS MEASUREMENT SUMMARIES

| System and Human Annotated Summaries | |
|---|---|
| Type | Summary |
| $S_{Summary}$ | 'Some of the genes in the BCAA metabolic pathway such as MLYCD (rank 164)HADHB (rank 354)IVD (rank 713)MUT (rank 921)and PCCB (rank 684) are also ranked highly by Hridaya. The SVMs are based on 181 features broadly grouped into (1) genetic(2) epigenetic(3) transcriptomic(4) phenotypicand (5) evolutionary. The genes are PDGFRBABL1FLT1; and these genes are drug targets of cancer drugs like Dasatinib (targets – PDGFRBABL1)Pazopanib (targets – PDGFRBFLT1)Ponatinib (target – ABL1)26.' |
| $H_{Model_1}$ | 'Some BCAA genes such as MLYCD, IVD , MUT and PCCB are ranked highly by Hridaya using SVM that is based on 181 features. These genes are drug targets of cancer drugs such as Dasatinib, Pazopanib and Ponatinib.' |
| $H_{Model_2}$ | 'A few genes in the BCAA metabolic pathway are also ranked highly by Hridaya and some examples include MUT (rank 921), IVD (rank 713), PCCB (rank 684), HADHE (rank 354) and MLYCD (rank 164). The SVMs are grouped into five categories based on 181 features and the categories are; genetic, epigenetic, transcriptomic, phenotypicand and evolutionary. The genes are PDGFRBABL1FLT1 and are drug targets of cancer drugs such as Dasatinib (targets – PDGFRBABL1), Pazopanib (targets – PDGFRBFLT1) and Ponatinib (target – ABL1)26.' |

Our result revealed that ROUGE-1 expressed better average result for the Recall (R), Precision (P), F1 score respectively with a 95% confident interval (see Table IV). The result revealed better evaluation metric in the *Recall* column of the *ROUGE* evaluation metrics. The results expressed better in $ROUGE{-}1$ with the *Recall* slightly over 83%, *Precision* slightly over 85% and $F1{-}Score$ slightly over 84% as reviewed in Table V.

Comparing the system generated summary with a new human

TABLE IV

ROUGE METRICS MEASUREMENT & ANALYSIS [S*summary*&*Hmodel*1]

| Average ROUGE Metrics | | | | |
|---|---|---|---|---|
| ROUGE | Recall | Precision | F1 Score | Conf.int |
| ROUGE-1 | 0.83784 | 0.40260 | 0.54386 | 95% |
| ROUGE-2 | 0.44444 | 0.21053 | 0.28572 | 95% |
| ROUGE-3 | 0.31429 | 0.14667 | 0.20000 | 95% |
| ROUGE-4 | 0.20588 | 0.09459 | 0.12962 | 95% |
| ROUGE-L | 0.78378 | 0.37662 | 0.50877 | 95% |
| ROUGE-W-1.2 | 0.34210 | 0.29676 | 0.31782 | 95% |
| ROUGE-S* | 0.69069 | 0.15721 | 0.25612 | 95% |
| ROUGE-SU* | 0.69943 | 0.16356 | 0.26512 | 95% |

summary model, produce a more appealing result. This was because the second summary was closely aligned with the original automated system summary (see Table V). This shows that the closeness of the human model summary to the system or reference summary produces better average across all ROUGE measuring dimensions (Recall, Precision and F1 score). Table V revealed a better and well-expressed precision results within all the *ROUGE* metrics used for the study evaluation. *ROUGE*− 1 shows the best percentage measure.

TABLE V

ROUGE METRICS MEASUREMENT & ANALYSIS [S*summary*&*Hmodel*2]

| Average ROUGE Metrics | | | | |
|---|---|---|---|---|
| ROUGE | Recall | Precision | F1 Score | Conf.int |
| ROUGE-1 | 0.83544 | 0.85714 | 0.84615 | 95% |
| ROUGE-2 | 0.56410 | 0.57895 | 0.57143 | 95% |
| ROUGE-3 | 0.37662 | 0.38667 | 0.38158 | 95% |
| ROUGE-4 | 0.22368 | 0.22973 | 0.22666 | 95% |
| ROUGE-L | 0.60759 | 0.62338 | 0.61538 | 95% |
| ROUGE-W-1.2 | 0.17792 | 0.43741 | 0.25295 | 95% |
| ROUGE-S* | 0.61960 | 0.65243 | 0.63559 | 95% |
| ROUGE-SU* | 0.62488 | 0.65756 | 0.64080 | 95% |

### *A. Procedure: Recall & Precision*

We have multiple processed articles or documents extracted from the web based on key search terms. The documents are stored in a given name *CleanHTML.txt* file and an automatic summary was generated and stored in a file called *summary.txt*. We then produced a set of human annotated reference summaries of the *CleanHTML.txt* document. The Recall in the context of the ROUGE metric simply means we are calculating how much of reference summary (the human summary) is the system summary (automated machine summary) recovering or capturing from our text. In considering the individual words in a sentence we simply represent this with the formula in equation 8.

$$ROUGE_{recal} = \sum_{match} \frac{count(overlapping_w)}{count(total_{ref}summary)} \quad (8)$$

The metric will produce a perfect result of 1 which usually will be the case if indeed the sentence matches. This metric simply means all the words in the reference summary has been captured by the system summary.

In the system generated summary, which sometimes might be very large based on the threshold selected, capturing all the words in the reference or model summary. However, most of the worlds in the system summary might be unnecessary verbose. But, this where precision becomes very important. In conducting precision on the summary, we are essentially measuring how much of the system or machine summary is required? We can measure precision using the equation 9.

$$ROUGE_{precision} = \sum_{match} \frac{count(overlapping_w)}{count(total_{sys}summary)} \quad (9)$$

This means we will evaluate and calculate words in the sentence summary of the *Recall* overlapping with the total words in the system summary. This will predict the words that are relevant which appears in the reference and over the total words in the system summary.The system's summary mostly contains unnecessary words in the summary. Therefore, our precision becomes crucial as we are trying to predict generated summaries that should be concise in nature. In this study, we combined and computerised both the *Precision* and *Recall* and further report the $F1 - score$ measure.

In order to ascertain the validity of the study, we measured ROUGE-N, $ROUGE - S$ and $ROUGE - L$ which are the granularity of texts that was compared between the system summaries and the reference or human annotated model summaries. $ROUGE - 1$ refers to the overlap of *unigrams* between the system summary and reference summary. $ROUGE - 2$ refers to the overlap of *bigrams* between the system reference and the model or reference summaries. We computed precision and recall scores of the $ROUGE - 2$. The main reason why $ROUGE-1$ could be considered over others or in conjunction with $ROUGE - 2$ or even other fine granularity measures is because it reveals the fluency of the summaries or if used in a translation task. The intuition is that following the word ordering of the reference m=summary indicate that the summary is more fluent.

The precision result tells us about the percentage (%) of the overlap between the system summary bigrams and the reference summary. We noticed in the case of the abstractive summarization as both the summaries of system and reference summaries get larger. There are few overlapping *bigrams* outcome as we are not always or directly re-using the whole sentences for the summarization.

## VII. RESULTS & FINDINGS

The terms in the topic modelling show text which are mostly frequent in the document these were depicted by the size of the circle (as seen in Figures 7 and 8 ). Representation of the result using scatter plot would reveal the distance between topics, the distribution and relationship between topic levels.

The distance between two or more topics is an approximation of their semantic relationship. Note that close topics such as topics 1, 2 and 3 are semantically related which describes the terms in the topics. As observed in Figures 7 and 8 the terms gene and disease are described in the articles in relation to the topics distribution. This reveals that topics 1, 2 and 3 are semantically distributed and have relationship on topic levels. These reveal five selected topics from the topic model analysed using the LDA model. The LDA model was one of the input argument together with the corpus and dictionary of the emerging terms used for the topic modelling. The slider ($\lambda$) in the web-based interactive visualization depicts the relevance metric of the rank terms. It is worth knowing that the terms of the topic are ranked in decreasing order by default in accordance with the topic-specific probability ($\lambda$ = 1). Figure 7 reveals the common terms from the topic model when the slider is at the full probability.

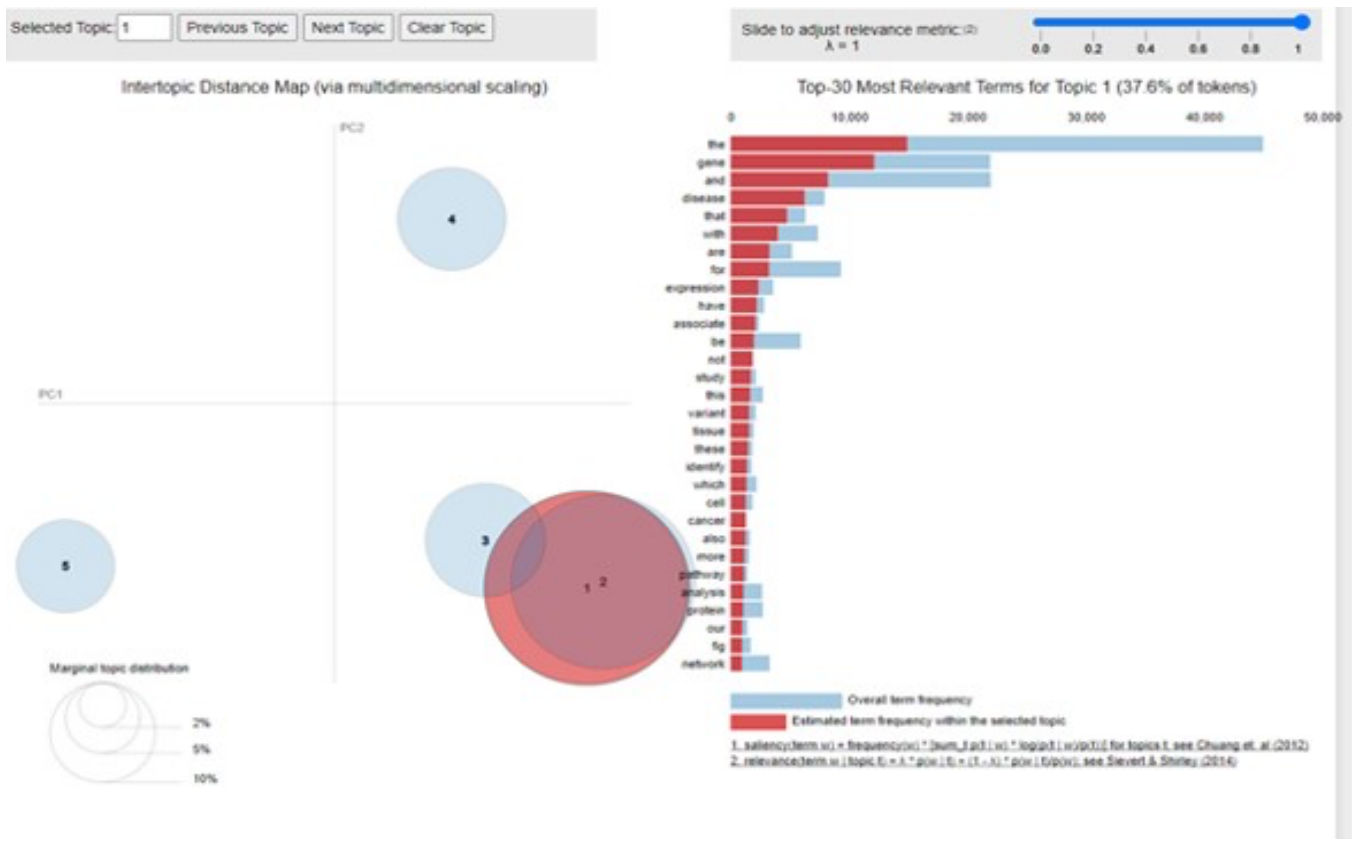

Fig. 7. Topic model visualization of interactive web-based topics

Note that the search key *gene* and *disease* were used to extract the text data (100 journal papers) from PubMed journal database related to the terms [22]. Figure 8 shows the most common terms in topic 2 to be *'gene'* and *'disease'* when the slider ($\lambda$ = 0.48) is positioned at 0.48 probability.

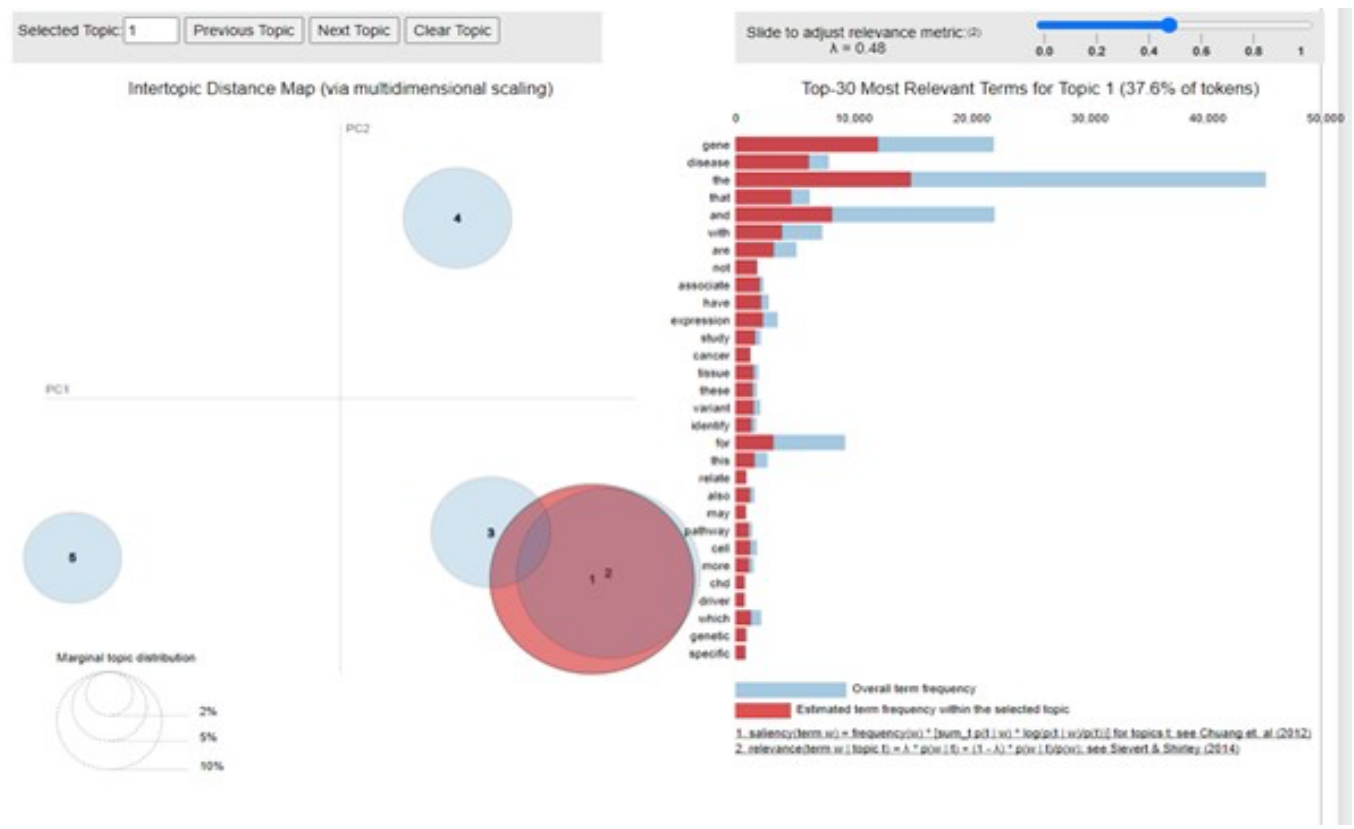

Fig. 8. Topic model visualization of interactive model show search terms predominantly projected

## VIII. Limitation

This study's limitations are observed in the precise summary prediction of articles of varying written styles. Some of the summarization models in most cases prefer nouns. Themes emerging from articles influences the grammatical structure of certain article summaries. Another limitation of the study was that the model takes longer to evaluate the emerging terms within the topic modelling approach used due to the large text data for analysis.

## IX. Discussion

In this study, we presented a fully data-driven approach for automatic text summarization. We proposed and evaluated the model on unstructured datasets which show some results comparable to the current state-of-the-art topic modelling techniques without depending on modifications using any linguistic information models [34]. Manual summarization is laborious and challenging task to accomplish. Therefore, automatization of the task is very essential. This process is gaining popularity among researchers. Summarization technique has been applied to various natural language processing (NLP) task such as in the areas of text analysis, classification, automated question and answering, financial and legal texts summarization, news summarization and reviewing of news headlines and the generation of social media articles [8], [35]. Performing research in these various topics could benefit from the early stages of document summaries which can be integrated into any base model at intermediate stages to help reduce the length of the document for further analysis.

## X. Conclusion

Automatic summarization is the process of reducing a text document with a computer program in order to create a summary that retains the most important points of the original document [36]. Text summarization is a very laborious problem to work on without accuracy in the summaries extracted from the documents. This study proposed fully automated single and multiple documents text summarization. Multiple documents were extracted and summarised while preserving the overarching meaning and purpose of the collective articles. The LDA model was one of the input arguments together with the corpus and dictionary of the terms that were used to perform the topic modelling in the study. The model designed within the study could conduct a cross-language text summarization where articles from other foreign languages could be processed and the summary translated into English and other languages. Our proposed future study will look into performing topic modelling with these documents and observe whether the approach retain the meaning of the original documents. The result from the future research will be compared with a current machine learning gene prediction application model designed for a new study on genes and diseases.